\documentclass[aps,twocolumn,showpacs,showkeys,prb,footinbib,superscriptaddress]{revtex4}
\usepackage{hyperref}  
\usepackage{graphicx}
\usepackage{dcolumn}
\usepackage{amsmath}
\begin{document}

\title{Decomposing the scattered field of two-dimensional metaatoms into multipole \textbf{contributions}}
\date{\today}



\author{J. Petschulat}
\email{joerg.petschulat@uni-jena.de}
\affiliation{Institute of Applied Physics, Friedrich-Schiller-Universit{\"a}t Jena,
Max Wien Platz 1, 07743, Jena, Germany}
\author{J. Yang}
\affiliation{Laboratoire Charles Fabry de l'Institut d'Optique, Universit\`{e}
Paris-Sud, 91127 Palaiseau cedex, France}
\author{C. Menzel}
\affiliation{Institute of Condensed Matter Theory and Solid State Optics,
Friedrich-Schiller-Universit{\"a}t Jena, Max Wien Platz 1, 07743, Jena, Germany}
\author{C. Rockstuhl}
\affiliation{Institute of Condensed Matter Theory and Solid State Optics,
Friedrich-Schiller-Universit{\"a}t Jena, Max Wien Platz 1, 07743, Jena, Germany}
\author{A. Chipouline}
\affiliation{Institute of Applied Physics, Friedrich-Schiller-Universit{\"a}t Jena,
Max Wien Platz 1, 07743, Jena, Germany}
\author{P. Lalanne}
\affiliation{Laboratoire Charles Fabry de l'Institut d'Optique, Universit\`{e}
Paris-Sud, 91127 Palaiseau cedex, France}
\author{A. T{\"u}nnermann}
\affiliation{Institute of Applied Physics, Friedrich-Schiller-Universit{\"a}t Jena,
Max Wien Platz 1, 07743, Jena, Germany}
\altaffiliation[also with: ]{Fraunhofer Institute of Applied Optics and Precision Engineering Jena, Germany.} 
\author{F. Lederer}
\affiliation{Institute of Condensed Matter Theory and Solid State Optics,
Friedrich-Schiller-Universit{\"a}t Jena, Max Wien Platz 1, 07743, Jena, Germany}
\author{T. Pertsch}
\affiliation{Institute of Applied Physics, Friedrich-Schiller-Universit{\"a}t Jena,
Max Wien Platz 1, 07743, Jena, Germany}

\begin{abstract}
We introduce a technique to decompose the scattered near field of two-dimensional
arbitrary metaatoms into its multipole contributions. To this end we
expand the scattered field upon plane wave illumination into cylindrical
harmonics as known from Mie's theory. By relating these cylindrical
harmonics to the field radiated by Cartesian multipoles, the contribution of the
lowest order electric and magnetic multipoles  can be identified.
Revealing these multipoles is essential for the design of metamaterials
because they largely determine the character of light propagation. In particular,
having this information at hand it is straightforward  to distinguish between
effects that result either from the arrangement of the metaatoms or from
their particular design.
\end{abstract}
 \maketitle

\section{Introduction}

Metamaterials may be understood as a kind of artificial matter that
allows to control the mould of light predominantly by the geometry of
their building blocks rather by their intrinsic material properties.
Fascination arose since these building blocks, commonly called the metaatoms,
can be designed to allow for propagation effects inaccessible in natural
materials. To simply describe the optical action of metamaterials,
effective properties are in most cases assigned that are retrieved from
the optical response of an ensemble of metaatoms instead from these
individual atoms themselves. For this purpose, single layer
\cite{Shen2009} or bulk metamaterials \cite{Zhang2008,Giessen2008} are
usually treated as black boxes to which effective properties are assigned with the
only purpose to reproduce scattering data like reflection and transmission
coefficients. These data do not provide sufficient insights into the physics
of metamaterials since their rational design usually aims at evoking a certain
multipolar scattering response \cite{Cho2008,Petschulat2008}. Although
metamaterials cover a wide range of structures at present, media with a
magnetic response \cite{Burresi2009,Decker2009} at optical frequencies are
particularly appealing since they do not exist in nature. The pertinent
metaatoms shall than possess a strong magnetic dipole moment. To achieve this,
the metaatoms are typically ring-shaped \cite{Zheludev2009}, resulting in a
ring-like current distribution at resonance. Hence, the optical response
contains a strong magnetic dipole field contribution leading to an appreciable
dispersion in the effective permeability. Such understanding of metamaterials is
very versatile as it provides the possibility to optimize metaatoms for different
spectral domains.

The advantage of understanding the optical response in terms of
multipole scattering is furthermore proven by various theoretical works
\cite{Cho2008,Petschulat2008,Petschulat2009,Moloney2009}. It was shown
that the optical response of single metamaterial layers as well as of bulk
metamaterials can be described by assuming induced multipole moment
densities up to the second order. From the multipolar contributions of the field
scattered by the metaatoms it is even possible to directly assign effective
material parameters. Even the effects of disorder can be studied and understood
in terms of a multipole analysis as shown theoretically and experimentally in Ref.
\onlinecite{Helgert2009b}. Hence, a detailed quantitative theoretical study of the
optical response of the single metaatom is in order and in most cases sufficient
to deduce the optical response of an ensemble of these entities. Although
important, such a contribution is currently missing. Whereas first
attempts are reported in literature \cite{Moloney2009} the analysis was usually
restricted to the far-field scattering response. Then, either by optimizing the
magnitude of the different multipolar contributions to match a certain angular
scattering response or by probing for the scattering strength in certain directions
where some multipole moments do not radiate, the multipolar response
can be revealed. Nevertheless, it remains an open question how
unique the assignments based on the far fields are.\\

In this contribution we develop a rigorous method to analyze the
scattered near-field of individual metaatoms that permits disclosing
their multipolar scattering contributions. The key ingredient is an expansion of
the scattered field of the metaatoms upon plane wave illumination into
cylindrical harmonics, i.e., we are restricting the current analysis to
two-dimensional metaatoms. By relating these cylindrical harmonics to
the field of Cartesian multipoles, it is possible to calculate their spectrally
resolved amplitudes. With this method at hand we will subsequently investigate
the multipole contributions to the scattered field of two prominent and
frequently studied metaatoms providing artificial magnetism; namely the
split-ring resonator (SRR) and the cut-wire pair (CW). It is shown that the
scattering response contains contributions of electric and magnetic dipoles, but
also of an electric quadrupole.  It resonates simultaneously with the
magnetic dipole and its contribution is much stronger for the CW when
compared to the SRR.

Although we are focussing here only on the analysis of two specific
metaatoms, the present technique is general and can be applied to various other
metaatoms as well. Perspectively it will permit to design
metaatoms with specific predefined multipolar contributions to
the scattered field and represent a tool to distinguish between
properties emerging from the periodic arrangement of
metaatoms  or from the specific metaatom scattering  response. This
will be of particular importance for the prediction of effective properties of
self-organized, bottom-up metamaterials which might not allow for a perfectly
periodic metaatom arrangement.

\section{Multipole expansion of the two-dimensional scattered field}
In order to reveal the multipolar character of the field scattered by an arbitrary
shaped metaatom, we will develop a method to expand its scattered field into
multipole fields. Since we focus here on two-dimensional structures, cylindrical
harmonics are an appropriate system of eigenfunctions. As soon as we have the
corresponding expansion at hand, we will show the equivalence between these
eigenfunctions and the Cartesian multipole fields for cylindrical sources. Hence,
this expansion will
allow for a direct calculation of the desired multipole coefficients.\\
We will start by briefly deriving  the mathematical background to decompose the
scattered field of a single metaatom into multipole fields. As known from Mie's
theory \cite{Mie1908} the translational invariance in $z$-direction allows to
separate the general vectorial scattering problem into the two scalar
cases of TE- and TM-polarization. For light propagating in a linear, homogenous,
isotropic, local medium the tangential fields satisfy the scalar wave equation
\cite{Jackson1975}
\begin{eqnarray}
\nabla_{x,y}^2{F}_z(x,y)+k(\omega)^2F_z(x,y)=0
\label{eq1}
\end{eqnarray}
with $k(\omega)^2=\omega^2/
c^2$. In Eq. (\ref{eq1}) $F_z(x,y)$ denotes the tangential
component of either the magnetic field for TM polarization (magnetic
component out-of-plane) or the electric field for TE polarization (electric field
component out-of-plane).

Casting Eq. (\ref{eq1}) into polar coordinates with
$x=R\cos(\phi),~y=R\sin(\phi),~R=\sqrt{x^2+y^2}$ we obtain [$F_z=F_z(R,\phi)$]
\begin{eqnarray}
\frac{1}{R}\frac{\partial}{\partial R}\left(R\frac{\partial }{\partial R}F_z\right)+\frac{1}{R^2}\left(R\frac{\partial^2}{\partial\phi^2}F_z\right)+k^2F_z=0.
\label{eq2}
\end{eqnarray}
The general solutions to this equation are radially dependent Bessel
functions with an azimuthally varying phase
\begin{eqnarray}
F_z=\sum_{m=-\infty}^{\infty}Z_m(kR)e^{im\phi}.
\label{eq3}
\end{eqnarray}
Eq. (\ref{eq3}) corresponds to the linearly independent solutions of
Eq. (\ref{eq2}) which consist of combinations of Bessel functions of the first $[J_m(kR)]$ and the second
kind $[Y_m(kR)]$ denoted as  $Z_m$ multiplied by an angular function.  Finally the scattered field from any cylindrical object centered
at the origin and subject to the Sommerfeld radiation condition at
infinity reads as
\begin{eqnarray}
F_{z,s}=\sum_{m=-\infty}^\infty a_m i^mH^{(1)}_m(kR)e^{im\phi},
\label{eq4}
\end{eqnarray}
where $H^{(1)}_m(kR)=J_m(kR)+iY_m(kR)$ are Hankel functions of the first kind
and $a_m$ are the expansion coefficients, termed Mie scattering
coefficients of the respective expansion order $m$. Note that any field outside a
virtual cylinder that entirely contains the scattering object can be expressed by
Eq. (\ref{eq4}) since these eigensolutions form an orthogonal and complete set of
eigenfunctions. By expanding the field scattered by an arbitrary particle in this
base we can determine the contribution of the respective expansion order $m$
to the total scattered field.

\begin{figure*}[]
\includegraphics[width=13cm]{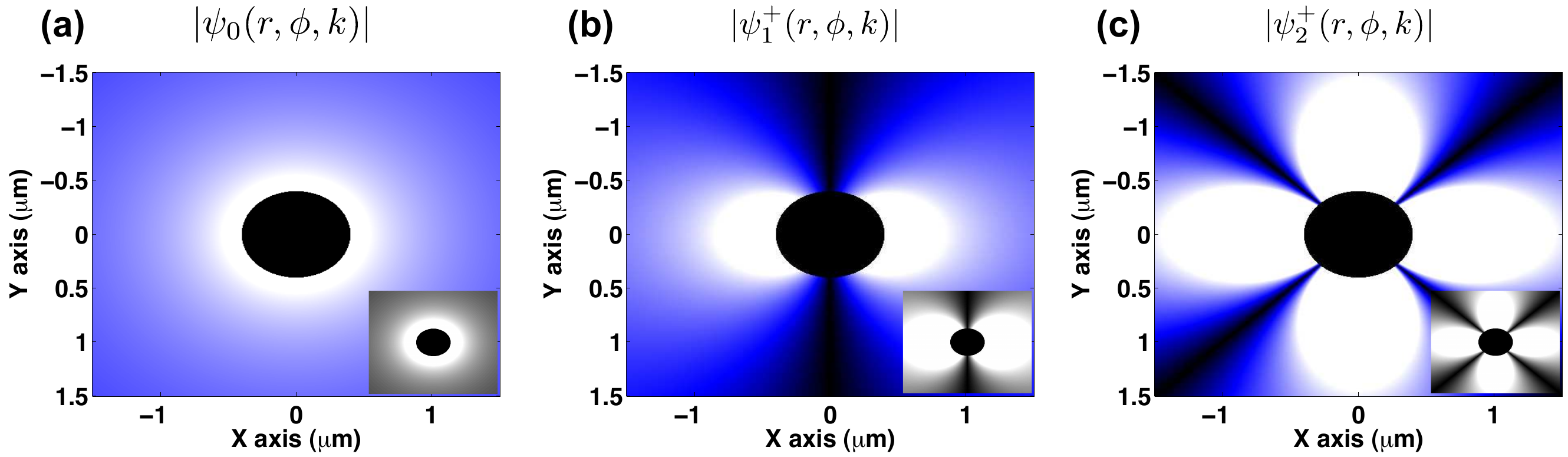}
\caption{(Color online) The modulus of the magnetic field distribution
(TM polarization) described by the eigenfunctions according to Eq. (\ref{eq5})
for the first three expansion orders. We have plotted the eigenfunctions
accounting for the $\cos(m\phi)$ angular distributions only. The results for the
$\sin(m\phi)$ terms would result in a rotation by an angle of $\pi/2m$. By
comparison with the field distributions of point multipoles, which are shown in
the gray-scaled insets, it can already be anticipated that the eigenfunctions
$\psi$ can be identified with multipole fields. Here $m=0$ (a) corresponds to a
magnetic dipole, $m=1$ (b)  is related to the electric dipole, while the electric
quadrupole is associated with $m=2$ (c).} \label{fig1}
\end{figure*}

The individual scattering order $m$  always consists of two parts, namely the
contributions from $m$ and $-m$ in Eq. (\ref{eq4}). Hence the scattered field
($F_{z,s}$) summarizing these orders can be rewritten \begin{eqnarray}
F_{z,s}&=&\sum_{m=0}^\infty \left[a_m^+\psi_m^+(R,\phi,k)+a_m^-\psi_m^-(R,\phi,k)\right], \nonumber \\
\psi_m^+(R,\phi,k)&=& i^mH_m^{(1)}(kR)\cos(m\phi),~ \nonumber\\
\psi_m^-(R,\phi,k)&=& i^{m+1}H_m^{(1)}(kR)\sin(m\phi), \nonumber \\
a^\pm_m&=&(a_m\pm a_{-m}), \label{eq5}
\end{eqnarray}
where we used $[H_m^{(1)}=H_{-m}^{(1)}]$. Now Eq. (\ref{eq5}) can be
considered as an usual series expansion with respect to
eigenfunctions of increasing order. These eigenfunctions
$\psi^\pm_m(R,\phi,k)$ are simply products of a radially and an
azimuthally varying function. Regarding the azimuthal terms it can be seen from
Eq. (\ref{eq5}) that they split into two contributions that are  $\pi/2m$
phase-shifted indicated by the superscripts $\pm$, respectively. This is related
to the fact that each expansion order is composed of two physically
identical but azimuthally rotated contributions of order $m$. For realistic
metaatoms the fundamental azimuthal part can be selected regarding
the symmetries of the nanostructure; in general both linear independent
contributions have to
be considered for the respective order $m$.\\

In order to decompose the scattered field of an arbitrary metaatom into the derived
set of eigenfunctions [Eqs.(\ref{eq5})] the Mie coefficients $a_m$ are
required. As usual they are obtained by evaluating the overlap integral between
the eigenfunction $\psi_m^\pm(R,\phi,k)$ and the
respective field component $F_z(R,\phi)$ of the individual scattering object\\
\begin{eqnarray}
a_m^\pm=\frac{\int_0^{2\pi}d\phi\int_{R_1}^{R_2}dRRF_z(R,\phi)\psi^{\pm\dagger}_m(R,\phi,k)}{\int_0^{2\pi}d\phi\int_{R_1}^{R_2}dRR|\psi^\pm_m(R,\phi,k)|^2}.
\label{eq10}
\end{eqnarray}
With Eq. (\ref{eq10}) it is possible to rigorously determine $a^\pm_m$ based on
the field overlap calculated on an annulus  with the two radii $R_2$ and
$R_1$. Here the orthogonality of the eigenfunctions $\psi^{\pm}_m(R,\phi,k)$
was exploited, i.e. replacing $F_z(R,\phi)$ by any eigenfunction of the order $l$
such that $F_z(R,\phi)=\psi_l(R,\phi,k)$ yields
\begin{widetext}
\begin{eqnarray}
%
\frac{\int_0^{2\pi}d\phi\int_{R_1}^{R_2}dRR\psi^\pm_l(R,\phi,k)\psi^{\pm\dagger}_m
(R,\phi,k)}{\int_0^{2\pi}d\phi\int_{R_1}^{R_2}dRR|\psi_m(R,\phi,k)|^2}
&=&\frac{\int_{R_1}^{R_2}dRRH^{(1)}_l(R,k)H^{(1)}_m(R,k)}{\int_0^{2\pi}d\phi\int_{R_1}^{R_2}dRR|\psi_m(R,\phi,k)|^2}\int_0^{2\pi}d\phi\left[\sin(l\phi)\sin(m\phi)+\cos(l\phi)\cos(m\phi)\right], \nonumber\\
&=&\frac{\int_{R_1}^{R_2}dRRH^{(1)}_l(R,k)H^{(1)}_m(R,k)}{\int_{R_1}^{R_2}dRR|H^{(1)}_m(kR)|^2}\delta_{ml}=\delta_{ml}.
\label{eq11}
\end{eqnarray}
\end{widetext}
By exploiting the orthogonality of the azimuthally varying part of
$\psi^{\pm}_m(R,\phi,k)$, it suffices to evaluate the overlap integral
for a fixed value of $R$ rather than an annulus to obtain $a_m^\pm$
\cite{Jackson1975}. However, it turned out that for  numerically (or
potentially experimentally) determined scattered fields of realistic metaatoms, the annulus integration is more stable and it was hence retained. This is not a
numerical inaccuracy but it is rather attributed to the discrete mesh where the
numerical data is available. This mesh  is not aligned with a cylinder surrounding
the object and an appropriate interpolation is required. Although
a sufficiently fine grid improves the stability this issue can be circumvented if the
annulus integration is performed.

\begin{figure*}[]
\includegraphics[width=15cm]{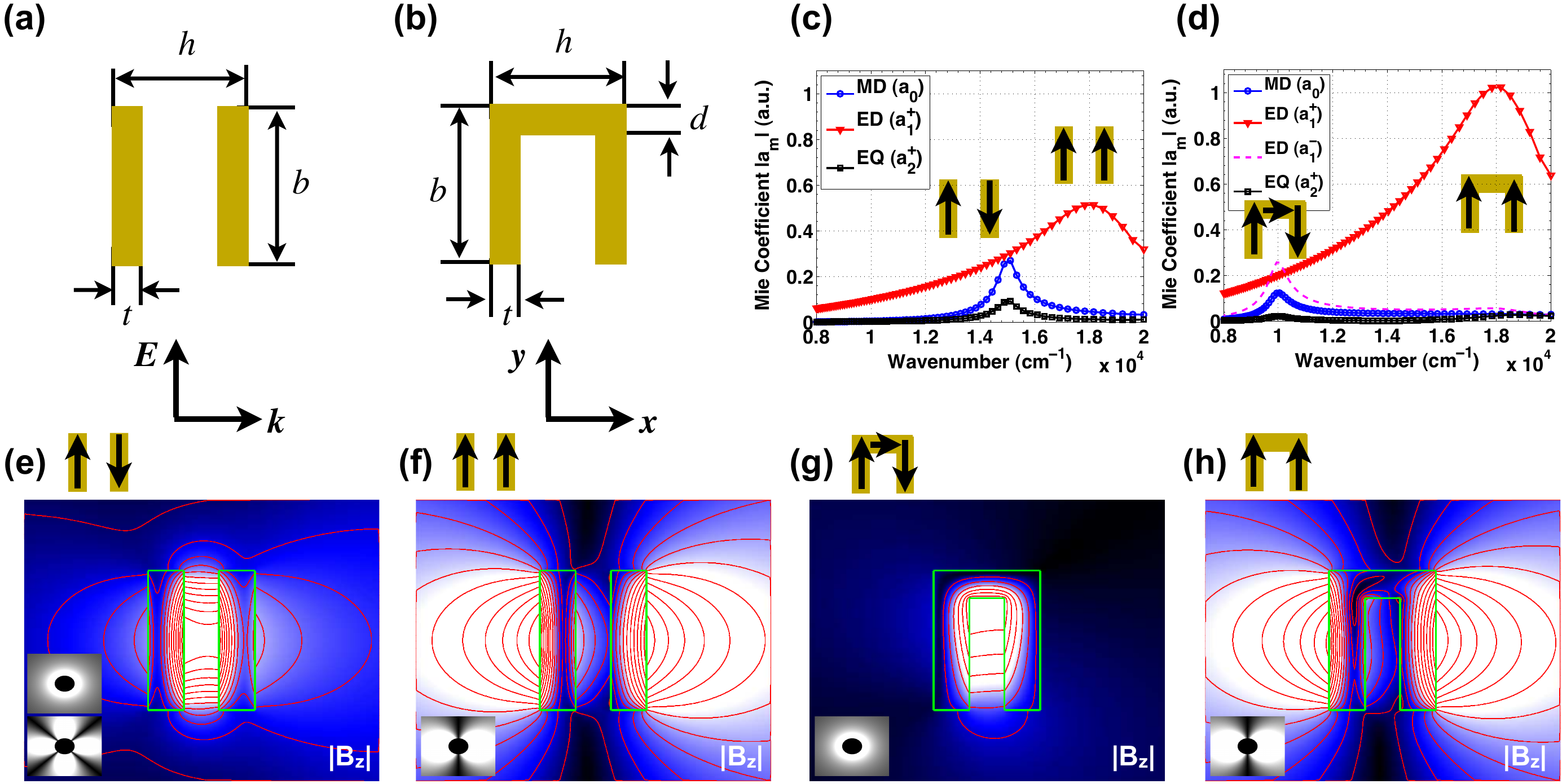}
\caption{(Color online) The illumination conditions, the orientation, and the
definition of the geometrical parameters of the CW (a) and the SRR metaatom
(b). The calculated Mie coefficients, related to the magnetic dipole (MD), the
electric dipole (ED) and the electric quadrupole (EQ) for the CW (c) and the SRR
(d) metaatom. Finally, the scattered magnetic fields for the {\it magnetic} (e) and
the {\it electric} resonance (f) of the CW and the SRR (g,h) are shown,
respectively. The gray-scaled insets show the exact multipole magnetic field
distributions as in Fig. \ref{fig1} to underline the similarities to the exact
scattering field patterns.} \label{fig2}
\end{figure*}

Although, it suffices to know the amplitudes $a_m^\pm$  to fully describe the
scattering response, they do not provide the physical insight as provided by
Cartesian multipole moments. Therefore, they need to be unambiguously
related. Since we have restricted ourselves to a two-dimensional configuration,
we have to relate the scattered field of a line source exhibiting any of the
relevant electric or magnetic multipole moments to the respective cylindrical
eigenfunctions. Starting with the well-known expressions for electrodynamic
point multipoles in Cartesian coordinates \cite{Jackson1975,Raab2005} we
derive below how they are related to $\psi_m(R,\phi,k)$. This is performed for
the three lowest orders by integrating the point multipoles in Cartesian
coordinates along the $z$-direction to disclose the radiation pattern of a line
source exhibiting such point multipoles.

The vector potential of monochromatic electromagnetic fields
originating from localized oscillating currents for the two lowest orders of a
multipole expansion reads as
\begin{eqnarray}
\mathbf{A}(\mathbf{r})  &=&\frac{\mu_0}{4\pi} \frac{e^{ikr}}{r}\int d^3r' \mathbf{j}(\mathbf{r}') \nonumber \\
&+&\frac{\mu_0}{4\pi} \frac{e^{ikr}}{r}\left(\frac{1}{r}-ik\right)\int d^3r' \mathbf{j}(\mathbf{r}')(\mathbf{n}\cdot\mathbf{r}'). \label{eq6}
\end{eqnarray}
The first term in Eq. (\ref{eq6}) accounts for the electric dipole moment where
$r$ is the length of the three dimensional radius vector
$(r=\sqrt{x^2+y^2+z^2})$ and $\mathbf{n}=\mathbf{r}/r$ is the normal vector.
The second term consists of the electric quadrupole and the magnetic dipole
contributions, both representing second-order moments in the multipole
expansion. After some algebra the electric dipole term can be written as
\cite{Jackson1975}
\begin{eqnarray}
\mathbf{A}_\mathrm{ed}(\mathbf{r})  &=&-i\omega\frac{\mu_0}{4\pi}\frac{e^{ikr}}{r}\mathbf{p}(\mathbf{r}). \label{eq7}
\end{eqnarray}
with $\mathbf{p}(\mathbf{r})=\int d^3r'\mathbf{r}'\rho(\mathbf{r}')$ being the
electric diploe moment. For electric dipole moments localized in the $(x,y)$ plane
the $z$ component of the magnetic field can be calculated as
\begin{eqnarray}
\mathbf{B}(\mathbf{r})&=&\nabla\times\mathbf{A}(\mathbf{r}), \nonumber \\
B_\mathrm{ed,z}(\mathbf{r})&=&-i\omega\frac{\mu_0}{4\pi}\frac{e^{ikr}}{r}\left[p_y\left(iky-\frac{y}{r^2}\right)\right. \nonumber \\
&-&\left.p_x\left(ikx-\frac{x}{r^2}\right)\right]. \label{eq8}
\end{eqnarray}
Since we are interested in the two-dimensional representation of the fields we
integrate Eq. (\ref{eq8}) along the $z$ axis (see appendix for details)
\begin{eqnarray}
B_\mathrm{ed,z}^\mathrm{2D}(x,y)&=&\omega\frac{\mu_0k}{4}\left[\cos(\phi)p_x-\sin(\phi)p_y\right] H_1^{(1)}(kR), \nonumber \\
&=&H_1^{(1)}(kR)\left[a_1^+\cos(\phi)-a_1^-\sin(\phi)\right], \nonumber \\
a_1^\pm&\equiv&\frac{\omega\mu_0}{4}p_{x,y}. \label{eq9}
\end{eqnarray}
As it can be easily verified, this electric dipole field coincides with the field
in Eq. (\ref{eq5}) for $m=1$. Hence, the eigenfunction of the order
$m=1$ represents the electric dipole contribution. Likewise one
can show that the magnetic dipole and the electric quadrupole contribution correspond to $m=0$ and $m=2$, respectively. A detailed
derivation for the multipoles can be found in the appendix. Based on these
results we can conclude that simply by calculating the Mie coefficients
$a_m^\pm$ we have the multipole coefficients of the scattered field at hand.

In Fig. \ref{fig1} the magnetic field patterns for $m=0,1,2$ are shown. Obviously
they correspond to the known field distributions for the magnetic dipole
($m=0$), the electric dipole ($m=1$) and the electric quadrupole ($m=2$)  as is
clear by comparison with the field patterns of point multipoles (see insets in Fig.
\ref{fig1}).

Having finished the analytical treatment, we are going to apply the results to
exemplarily reveal  the multipole scattering contributions for two
prominent metaatoms in the following.

\section{Multipole scattering of Metaatoms}

For the application of the technique developed above, we selected the CW
 and the SRR geometry. We emphasize that this is not a necessary restriction
because arbitrary structures can be investigated too with the developed
formalism. We selected these two structures since several modified, more
complex metamaterials are composed out of these basic plasmonic entities
\cite{Giessen2009,Zhang2008}. The investigated metaatoms for  both
scenarios are shown in Fig.~\ref{fig2}(a,b). The CW structure has a wire distance
of $h=60~\text{nm}$, a wire thickness of $t=20~\text{nm}$ and a width of
$b=100~\text{nm}$. To evaluate the overlap integral
 the annulus radii were set to $R_1= 70~\text{nm}$ and $R_2=80~\text{nm}$. In order to observe the localized
eigenmodes at similar spectral positions we used the same dimensions for the
SRR with an additional connection ($d=20~\text{nm}$) of both wires [Fig. \ref{fig2}(b)]. As a material
for both metaatoms we selected gold \cite{JC1972} embedded in vacuum. The
structures were illuminated according to the
conditions as shown in Fig.~\ref{fig2}(a,b) with monochromatic plane waves.\\

In order to calculate the electromagnetic near fields we applied the finite element
method (FEM) \footnote{We applied the commercial product COMSOL.
(www.comsol.com)}. By calculating the scattering patterns for both structures
and performing the field overlap calculations according to Eq. (\ref{eq6}) for the
first three orders one obtains the Mie coefficients as shown in Fig. \ref{fig2}(c,d).
For both metaatoms the observed low energy resonance peaks occur for both second order multipole
contributions (electric quadrupole and magnetic dipole) at the same frequency
(wavenumber), while the high frequency resonances are associated with the
electric dipole modes. Thus, the usual argument that the electric dipole
contribution of the two currents, oscillating  $\pi$ out-of-phase in the
wires perpendicular to the propagation direction, annihilate and hence the next
higher order multipole moments prevail, is confirmed by these results.
Furthermore, the quantitative contribution of each multipole moment is clearly
revealed. For the CW structure [Fig. \ref{fig2}(c)] the electric quadrupole
contribution is much stronger than that for the SRR. This is due to the different
symmetry of both metaatoms. The shortcut of both wires essentially
prevents the excitation of a quadrupole moment. Hence, the SRR's optical
response is mainly governed by the electric and the magnetic dipole
moment.

This argumentation is supported by considering the near fields at resonance.
Comparing the pure multipole fields (Fig. \ref{fig1}) with the scattered fields of
the two metaatoms [Fig. \ref{fig2}(e-h)] it becomes obvious that the magnetic
field of the high frequency resonance (electric resonance) compares nicely to that of
an electric dipole. The scattered magnetic fields for the low frequency resonance
(magnetic resonance) for the CW structure shows a combination of
centro-symmetric magnetic dipole fields in between the wires and the electric
quadrupole fields around. This is indicated by the fourfold patterns in the
iso-surface lines in Fig. \ref{fig2}(e) outside the CW geometry. Differing from
these patterns the SRR exhibits near field features that manly attributed to
the magnetic dipole radiation pattern [Fig. \ref{fig2}(g)]. In addition a weak background contribution of an electric dipole tilted by $45^\circ$ can be observed in Fig. \ref{fig2}(g), caused by the superposition of weak SRR electric dipole moments in $x$ and $y$ directions [e.g. see Fig. \ref{fig2}(d)].
\\
Note that with the knowledge of the scattering response of the isolated
metaatoms other physical observables become easily accessible. Most notably
various other important metamaterial properties, i.e. the effective cross sections
\cite{Husnik2008} or the isolated polarizabilities  \cite{Celebrano2009} that can
be obtained by this Mie theory based formalism straightforwardly, can be
analytically calculated  \cite{Bohren1983}.

\begin{figure}
\includegraphics[width=8.5cm]{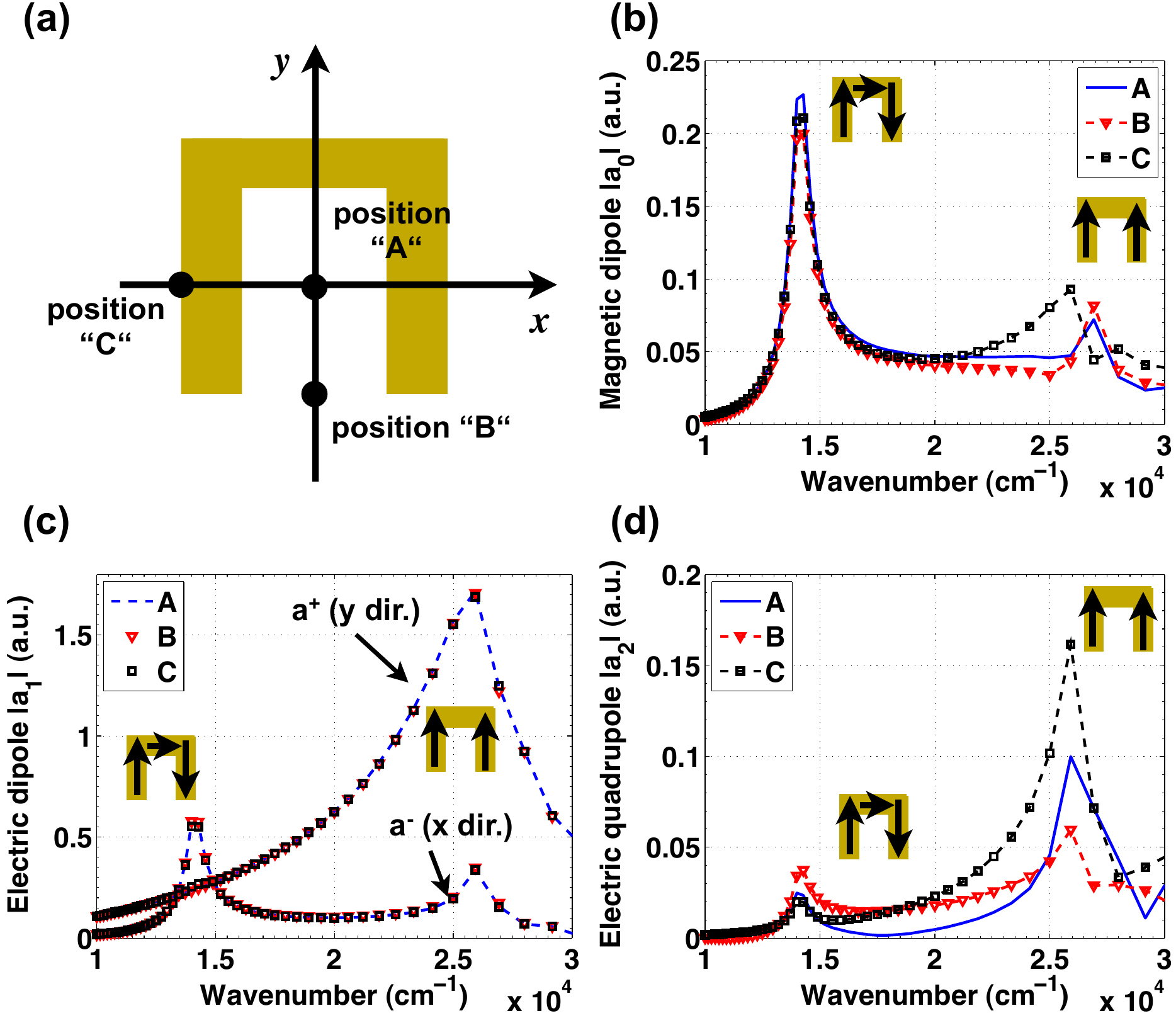}
\caption{(Color online) (a) The SRR geometry together with the three positions representing the selected origins for the multipole expansion.
 "A" is associated with the center of symmetry of the corresponding CW structure, while "B" and "C" represent additional positions out of the
 metaatoms center. The calculated Mie coefficients for  the magnetic dipole (MD)(b),the electric dipole (ED) (c) and the electric quadrupole (EQ) (d).
 For completeness both electric dipole moments along the $x$ $(a_1^+)$ and the $y$ axis $(a_1^-)$ are shown.}
\label{fig3}
\end{figure}

Finally, the dependence of the multipole coefficients on the choice of the origin is
investigated. From electrostatics it is known that the first non-vanishing
multipole moment does not depend on the choice of the origin. For optical fields
this is not valid anymore. However, for the scattering response of very small
nanoparticles where the quasi-static approximation is valid, this dependency is
expected to be negligible for reasonably chosen origins. Exemplarily the
multipole expansion is performed for the scattered field of a SRR with smaller
dimensions ($h=70~nm,~b=70~nm,~t=25~nm,~d=20~nm$) to prove the
quasi-statically expected behavior for three different origins. The corresponding
results are shown in Fig. \ref{fig3}. Clearly the electric dipole moments are
almost constant for different origins, whereas the magnetic dipole as well as the
electric quadrupole moments may appreciably deviate, in particular for the
second resonance. Of course, the multipole moments will strongly deviate  from
the results shown here if the origin is placed far outside the metaatoms, but then
any multipole expansion becomes meaningless, too. Physically, the origin should
be chosen such that all higher order multipoles beyond the second order are
strongly suppressed.

Nevertheless from these results we can conclude that for reasonably chosen
origins, i.e. placed close to the center of mass as suggested in Ref.
\onlinecite{Moloney2009} the multipolar character of the scattered field can be
revealed consistently. For calculating effective material parameters on the other
hand this origin dependence has to be kept in mind \cite{Raab2005}.

\section{Summary}
In summary, we presented a novel technique based on Mie
theory to reveal the contribution of various multipole excitations to the near-field
scattering pattern of metaatoms. For the sake simplicity we restricted ourselves to two-dimensional metaatoms. Since Mie theory has been originally
developed for three-dimensional objects (spheres) the approach can be also
extended towards three-dimensional metaatoms. With the presented formalism we
revealed the quantitative contributions of multipole moments to the
plasmonic eigenmodes of the cut-wire and the split-ring resonator structures. As
previously phenomenologically interpreted, due to the similarities of the multipole moments and the electron dynamics, we could rigorously confirm the excitation of up
to second order multipoles as the dominant scattering contributions. We
anticipate that such a multipole expansion of the scattered field is a genuine
approach to optimize metaatoms with regard to a predefined scattering
response. It will be very helpful to identify the effect of periodicity or its absence, in case of 
metamaterials fabricated with self-organization techniques, with respect to the
effective MM properties.
Beyond possible applications in the field of metamaterials, the presented approach may also provide guidelines in the design of optical nanoantennas to achieve a desired radiation
characteristics to be matched to an, in principle, arbitrary source.

\section{Acknowledgements}
Financial support by the Federal Ministry of Education and Research (MetaMat
and PhoNa), the State of Thuringia within the Pro-Excellence program (MeMa),
and the European Union (NANOGOLD) is acknowledged. Part of the work of JY
was supported by the Erasmus Mundus Master Program OpSciTech. The work
was furthermore supported by the DAAD and the French MESR within the
PROCOPE exchange program.

\appendix*
\section{The connection between 3D and 2D radiated multipole fields}
\subsection{Electric dipole}
We start our considerations with the vector potential for the first two expansion
orders Eq. (\ref{eq6})
\begin{eqnarray}
\mathbf{A}(\mathbf{r})  &=&\frac{\mu_0}{4\pi} \frac{e^{ikr}}{r}\int d^3r' \mathbf{j}(\mathbf{r}') \nonumber \\
& &+\frac{\mu_0}{4\pi} \frac{e^{ikr}}{r}\left(\frac{1}{r}-ik\right)\int d^3r' \mathbf{j}(\mathbf{r}')(\mathbf{n}\cdot\mathbf{r}'). \nonumber\\
\label{Aeq1}
\end{eqnarray}
At first we consider the first order term on the right hand side of Eq. (\ref{Aeq1})
which represents the electric dipole contribution. Applying the continuity equation and requiring that the current density vanishes at infinity,
we arrive at the well-known vector potential of an electric dipole
\begin{eqnarray}
\mathbf{A}_\text{ed}(\mathbf{r})=-i\omega\frac{\mu_0}{4\pi}\frac{e^{ikr}}{r}\mathbf{p}(\mathbf{r}). \label{Aeq2}
\end{eqnarray}
According to Maxwell's equations we obtain the $z$-component of the magnetic field 
by applying the curl operator to the potential [Eq. (\ref{Aeq2})]
\begin{eqnarray}
B_{\text{ed},z}(\mathbf{r})&=&\frac{\partial}{\partial x}A_y(\mathbf{r})-\frac{\partial}{\partial y}A_x(\mathbf{r}), \nonumber \\
&=&-i\omega\frac{\mu_0}{4\pi}\left(p_y\frac{\partial}{\partial x}-p_x\frac{\partial}{\partial y}\right) \frac{e^{ikr}}{r}. \label{Aeq3}
\end{eqnarray}
In order to obtain the associated two-dimensional fields we integrate the
three-dimensional fields along the $z$ axis. This is equivalent to the transition from
{\it point} multipoles to {\it line} multipoles, similar to the transition from the
three-dimensional  to the two-dimensions Green's function. \cite{Martin1998}
\begin{eqnarray}
B_{\text{ed},z}^{2\text{D}}(x,y)&=&\int_{-\infty}^\infty dz B_{\text{ed},z}, \nonumber \\
&=&-i\omega\frac{\mu_0}{4\pi}\left(p_y\frac{\partial}{\partial x}-p_x\frac{\partial}{\partial y}\right) \nonumber \\
& & \times\int_{-\infty}^\infty dz \frac{e^{ikr}}{r}. \label{Aeq4}
\end{eqnarray}
The integral can be carried out with the help of Eq. 8.421 in
\cite{Gradsteyn1981}
\begin{eqnarray}
\int_{-\infty}^\infty dz \frac{e^{ik\sqrt{R^2+z^2}}}{\sqrt{R^2+z^2}}=i\pi H_0^{(1)}(kR). \label{Aeq5}
\end{eqnarray}
Now the recursion formula for Hankel's functions of the first kind can be applied
to calculate the remaining derivatives
\begin{eqnarray}
\frac{\partial}{\partial z}H_n^{(1)}(z)&=&\frac{n}{z}H_n^{(1)}(z)-H_{n+1}^{(1)}(z), \nonumber \\
\frac{\partial}{\partial X_l}H_0^{(1)}(kR)&=&-H_1^{(1)}(kR)\frac{X_lk}{R},~ X_l\in\{x,y\}. \nonumber \\ \label{Aeq6}
\end{eqnarray}
which allows to calculate the required magnetic fields as
\begin{eqnarray}
B_{\text{ed},z}^{2\text{D}}(R,\phi)&=&\omega\frac{\mu_0 k}{4}\left[\cos(\phi)p_x-\sin(\phi)p_y\right] H_1^{(1)}(kR). \nonumber \\
\label{Aeq7}
\end{eqnarray}
According to Eq. (\ref{eq5}) the associated magnetic field for $m=1$  is given by
\begin{eqnarray}
F_{z,s}&=&a_1^+\psi_1^+(R,\phi,k)+a_1^-\psi_1^-(R,\phi,k), \nonumber \\
&=&\left[a_1^+\cos(\phi)-a_1^-\sin(\phi)\right] H_1^{(1)}(kR). \label{Aeq8}
\end{eqnarray}
It becomes obvious that the coefficients $a_1^\pm$ represent two electric
dipoles rotated by $\pi/2$, since a comparison of the coefficient  between Eq.
(\ref{Aeq7}) and Eq. (\ref{Aeq8}) yields
\begin{eqnarray}
a_1^+=\omega\frac{\mu_0 k}{4}p_x, \qquad a_1^-=\omega\frac{\mu_0 k}{4}p_y. \label{Aeq9}
\end{eqnarray}
\begin{figure*}
\includegraphics[width=17cm]{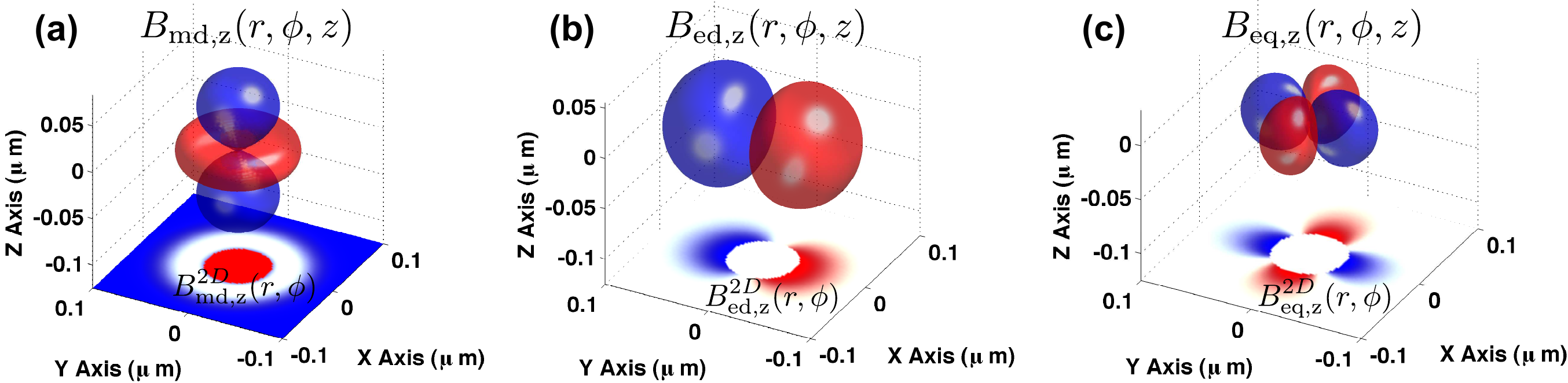}
\caption{(Color online) The exact field patterns for the $z$ component of the
magnetic field for  the magnetic dipole (a), the electric dipole (b), and  the
electric quadrupole (c) for carrier dynamics in the $(x,y)$ plane. The
two-dimensional field distributions below the three-dimensional ones show the
 three-dimensional fields for the respective multipole moment integrated along the $z$ axis.
The moduli of these two-dimensional fields correspond to the insets in
Fig. \ref{fig1}(a-c) and precisely agree with $\psi_m(r,\phi),~m\in\{1,2,3\}$,
respectively as shown in the derivation below.} \label{fig4}
\end{figure*}
\subsection{Magnetic dipole}
The second order expansion term in Eq. (\ref{Aeq1}) can be split into two
contributions where one is associated with the vector potential of  the electric
quadrupole and the other one with the magnetic dipole. Hence, both
quantities represent second-order moments
\begin{eqnarray}
\mathbf{A}(\mathbf{r})  &=&\frac{\mu_0}{4\pi} \frac{e^{ikr}}{r}\left(\frac{1}{r}-ik\right)\int d^3r' \mathbf{j}(\mathbf{r}')(\mathbf{n}\cdot\mathbf{r}') \nonumber \\
&=& \mathbf{A}_\text{md}(\mathbf{r})+\mathbf{A}_\text{eq}(\mathbf{r}), \nonumber \\
\mathbf{A}_\text{eq}(\mathbf{r})  &=&\frac{\mu_0}{4\pi} \frac{e^{ikr}}{r}\left(\frac{1}{r}-ik\right)\nonumber \\
& &\times\underbrace{\int_{-\infty}^\infty d^3r'\frac{1}{2}\left[(\mathbf{n}\cdot\mathbf{r}')\mathbf{j}(\mathbf{r}')+(\mathbf{n}
\cdot\mathbf{j}(\mathbf{r}'))\mathbf{r}'\right]}_{\propto\mathbf{Q}}, \nonumber \\
\mathbf{A}_\text{md}(\mathbf{r})  &=&\frac{\mu_0}{4\pi} \frac{e^{ikr}}{r}\left(\frac{1}{r}-ik\right)\nonumber \\
& &\times\underbrace{\left[\frac{1}{2}\int_{-\infty}^\infty d^3r'\mathbf{r}'\times\mathbf{j}(\mathbf{r}')\right]}_{\equiv\mathbf{m}}\times\mathbf{n}, \nonumber \\
&=&\frac{\mu_0}{4\pi} \frac{e^{ikr}}{r}\left(\frac{1}{r}-ik\right)\mathbf{m}\times\mathbf{n}.\label{Aeq10}
\end{eqnarray}
 Currents in the $(x,y)$
plane only induce a magnetic dipole moment $\mathbf{m}$ pointing into $z$ direction
\begin{eqnarray}
\mathbf{A}_\text{md}(\mathbf{r})&=&\frac{\mu_0}{4\pi} \frac{e^{ikr}}{r^2}\left(\frac{1}{r}-ik\right)
\left(-y\mathbf{e}_x+x\mathbf{e}_y\right)m_z. \nonumber \\
\label{Aeq11}
\end{eqnarray}
Using the identity
\begin{eqnarray}
\frac{\partial}{\partial X_j}\frac{e^{ikr}}{r}&=&-X_j\frac{e^{ikr}}{r^2}\left(\frac{1}{r}-ik\right),~ X_j\in\{x,y\}. \nonumber \\
\label{Aeq12}
\end{eqnarray}
Eq. (\ref{Aeq11}) can be simplified yielding
\begin{eqnarray}
\mathbf{A}_\text{md}(\mathbf{r})&=&\frac{\mu_0m_z}{4\pi}\left(\mathbf{e}_x\frac{\partial}{\partial y}-\mathbf{e}_y\frac{\partial}{\partial x}\right)\frac{e^{ikr}}{r}. \label{Aeq13}
\end{eqnarray}
In order to get the magnetic field Eq. (\ref{Aeq3}) can be applied which has to be
integrated along the $z$ axis as for the electric dipole before
\begin{eqnarray}
B_{\text{md},z}(\mathbf{r})&=&-\frac{\mu_0m_z}{4\pi}\left(\frac{\partial^2}{\partial x^2}+\frac{\partial^2}{\partial y^2}\right)\frac{e^{ikr}}{r}. \label{Aeq14}
\end{eqnarray}
Eq. (\ref{Aeq14}) can be easily integrated applying again Eq. (\ref{Aeq5})
\begin{eqnarray}
B_{\text{md},z}^{2\text{D}}(x,y)&=&i\frac{\mu_0\mu (\omega)m_z}{4}k^2\nonumber \\
& &\times\left[\frac{2}{kR}H_1^{(1)}(kR)-H_2^{(1)}(kR)\right], \nonumber \\ \label{Aeq15}
\end{eqnarray}
which can be simplified using the theorem 8.473, 3 in \cite{Gradsteyn1981}
\begin{eqnarray}
H_0^{(1,2)}(z)&=&\frac{2}{z}H_1^{(1,2)}(z)-H_2^{(1,2)}(z), \label{Aeq16}
\end{eqnarray}
resulting in the expected radially dependent expression for the scattered
magnetic field
\begin{eqnarray}
B_{\text{md},z}^{2\text{D}}(R,\phi)&=&i\frac{\mu_0m_zk^2}{4}H_0^{(1)}(kR). \label{Aeq17}
\end{eqnarray}
For comparison Eq. (\ref{eq5}) with $m=0$ yields
\begin{eqnarray}
F_{x,s}&=&a_0H_0^{(1)}(kR). \label{Aeq18}
\end{eqnarray}
Comparing Eq. (\ref{Aeq17}) and Eq. (\ref{Aeq18}) we see that the coefficient
$a_0$ again is directly proportional to the magnetic dipole moment, similar to
the electric dipole moment before [Eq. (\ref{Aeq9})]. Here the second term
connected to a rotated magnetic dipole by $\pi/2$ vanishes, as expected for the
radial symmetric radiation pattern of a magnetic dipole.

\subsection{Electric quadrupole}
The second term connected to the electric quadrupole moment in the vector
potential of Eq. (\ref{Aeq10}) can be simplified using Gauss' law and the
continuity equation to \cite{Jackson1975}
\begin{eqnarray}
\mathbf{A}_\text{eq}(\mathbf{r})  &=&\frac{\mu_0}{8\pi} \frac{e^{ikr}}{r}\left(\frac{1}{r}-ik\right)\nonumber \\
& &\times\int_{-\infty}^\infty d^3r'\mathbf{r}'(\mathbf{n}\cdot\mathbf{r}')\rho(\mathbf{r}'). \label{Aeq19}
\end{eqnarray}
In the following we restrict the carrier dynamics to the $(x,y)$ plane that
might be expressed by
\begin{eqnarray}
\rho(\mathbf{r}')=\sum_{\alpha=1}^Nq_\alpha\delta[x'-x_\alpha(t)]\delta[y'-y_\alpha(t)]\delta(z'). \label{Aeq20}
\end{eqnarray}

We now rewrite the integral kernel of Eq. (\ref{Aeq19})
\begin{eqnarray}
\int_{-\infty}^\infty d^3r'\mathbf{r}'(\mathbf{n}\cdot\mathbf{r}')\rho(\mathbf{r}')&=&\sum_{j=1}^3\mathbf{e}_j\sum_{l=1}^3\int_{-\infty}^\infty d^3r' X'_j n_lX_l'\rho(\mathbf{r}'), \nonumber \\
&\equiv& \hat{Q}_{jl}(\mathbf{n})\mathbf{e}_j. \label{Aeq21}
\end{eqnarray}
Here we have used $\hat{Q}(\mathbf{n})$ 
\begin{eqnarray}
\hat{Q}(\mathbf{n})&=&\sum_{\alpha=1}^Nq_\alpha\left(\begin{array}{ccc} x_\alpha^2n_x & x_\alpha y_\alpha n_y& 0\\ y_\alpha x_\alpha n_x & y_\alpha^2 n_y & 0\\ 0 & 0 & 0\end{array}\right), \nonumber \\
&\equiv&\left(\begin{array}{ccc} Q_{xx}n_x & Q_{xy} n_y& 0\\ Q_{xy} n_x & Q_{yy} n_y & 0\\ 0 & 0 & 0\end{array}\right), \label{Aeq22}
\end{eqnarray}
where the $Q_{ij}$ are the primitive symmetric quadrupole tensor
entries \cite{Raab2005}. Substituting Eq. (\ref{Aeq22}), (\ref{Aeq21}) into Eq.
(\ref{Aeq19}) and applying Eq. (\ref{Aeq12}) we obtain the vector potential
\begin{eqnarray}
\mathbf{A}_\text{eq}(\mathbf{r})  &=&-\frac{\mu_0}{8\pi}
\left[\mathbf{e}_x\left(Q_{xx}\frac{\partial}{\partial x}+Q_{xy}\frac{\partial}{\partial y}\right)\nonumber\right. \\
& &\left.+\mathbf{e}_y\left(Q_{yx}\frac{\partial}{\partial x}+Q_{yy}\frac{\partial}{\partial y}\right)\right]\frac{e^{ikr}}{r}. \label{Aeq23}
\end{eqnarray}
This equation can again be used to calculate the respective magnetic field [Eq. (\ref{Aeq3})]
\begin{eqnarray}
B_{\text{eq},z}(\mathbf{r}) &=& -\frac{\mu_0}{8\pi}\left[Q_{yx}\left(\frac{\partial^2}{\partial x^2}-\frac{\partial^2}{\partial y^2}\right)\right.\nonumber \\
& & \left.+Q_{yy}\frac{\partial^2}{\partial x\partial y}-Q_{xx}\frac{\partial}{\partial y\partial x}\right]\frac{e^{ikr}}{r}. \label{Aeq24}
\end{eqnarray}
 In a final stepEq. (\ref{Aeq14}) can be integrated by applying Eq. (\ref{Aeq5})
\begin{eqnarray}
B_{\text{eq},z}^{2\text{D}}(x,y)&=&-i\frac{\mu_0}{8}\left[Q_{xy}\left(\frac{\partial^2}{\partial x^2}-\frac{\partial^2}{\partial y^2}\right)\right.\nonumber\\
& &+\left.\left(Q_{yy}-Q_{xx}\right)\frac{\partial^2}{\partial x\partial y}\right] H_0^{(1)}(kR),\nonumber\\ \label{Aeq25}
\end{eqnarray}
which can be simplified using the recursion formula Eq. (\ref{Aeq6}) for the
second derivatives
\begin{eqnarray}
B_{\text{eq},z}^{2\text{D}}(R,\phi)&=&-i\frac{\mu_0k^2}{8}\left\{Q_{xy}[\cos^2(\phi)-\sin^2(\phi)] \right.\nonumber \\
& &+\left.(Q_{yy}-Q_{xx})\cos(\phi)\sin(\phi)\right\}H_2^{(1)}(kR). \nonumber \\ \label{Aeq26}
\end{eqnarray}
Finally we end up with the magnetic field
\begin{eqnarray}
B_{\text{eq},z}^{2\text{D}}(R,\phi)&=&-i\frac{\mu_0k^2}{16}[Q_{xy}\cos(2\phi)\nonumber \\
& &+\left(Q_{yy}-Q_{xx}\right)\sin(2\phi)]H_2^{(1)}(kR).  \nonumber\\
\label{Aeq27}
\end{eqnarray}
The evaluation of Eq. (\ref{eq5}) for $m=2$ yields
\begin{eqnarray}
F_{z,s}=-\left[a^+_2\cos(2\phi)+ia^-_2\sin(2\phi)\right]H_2^{(1)}(kR).  \nonumber \\
\label{Aeq28}
\end{eqnarray}
Again the coefficients $a_2^\pm$ are proportional to the electric quadrupole
moment entries under consideration.

\bibliography{PRBnearfields}
\end{document}